\documentstyle[twoside,fleqn,espcrc2]{article}

\newcommand{\AmS}{{\protect\the\textfont2
  A\kern-.1667em\lower.5ex\hbox{M}\kern-.125emS}}

\title{Monopole effects on Polyakov loop and its gauge independence 
       in QCD}

\author{Yoshimi Matsubara\address{Nanao Junior College,
         Nanao, Ishikawa 926, Japan}%
 \thanks{Presented by Y.Matsubara.},
        Sawut Ilyar\address{Department of Physics,
            Kanazawa University, Kanazawa 920-11, Japan},
        Tsuyoshi Okude$^{\small b}$,
        Kenji Yotsuji$^{\small b}$   
        and 
        Tsuneo Suzuki$^{\small b}$
}
       
\begin{document}

\begin{abstract}
Monte-Carlo simulations of abelian projection in 
$T \neq 0$ pure lattice QCD show that
1)\ Polyakov loops written 
in terms of abelian link fields alone play a role of an order parameter of 
deconfinement transition, 
2)\ the abelian Polyakov 
loops are decomposed into contributions from Dirac strings of monopoles 
and from photons, 
3)\ vanishing of the abelian Polyakov loops 
in the confinement 
phase is due to the Dirac strings alone and 
the photons give a finite contribution 
in both phases. 
Moreover, these results appear to hold good in unitary gauges. This
suggests that monopole condensation
as the color confinement mechanism is gauge independent.
\end{abstract}

\maketitle

\section{Introduction}

It is one of the most important subjects to understand mechanism of color
confinement in QCD.
The abelian projection of QCD is to extract an abelian theory 
with charges and monopoles performing  a partial gauge-fixing.  
 'tHooft conjectured an interesting idea \cite{thooft2}
 that condensation of the abelian monopoles produces color 
confinement due to the dual Meissner effect.

There are infinite ways of extracting an abelian theory out of QCD.
Maximally abelian(MA) gauge is exciting. 
In this gauge, many interesting facts have been found.
There are phenomena called abelian dominance \cite{suzu93}.
 The string tension from abelian Wilson loops is produced 
only by monopoles \cite{shiba,ejiri,stack}.
 The effective monopole action is derived from vacuum 
configulations \cite{shiba3}. In the infinite volume limit, entropy 
dominance over energy induces the condensation of the monopoles,
which leads us to the confinement of the abelian 
charges ,i.e., the color confinement.
 Therefore the (extended) monopoles control the confinement 
mechanism in MA gauge.

 The following subjects are to be clarified:

1) relation of the monopole dynamics to the deconfinement 
transition and

2) gauge dependence of the monopole dynamics.

 From this point of view, we investigate the Polyakov loop
which is an important order parameter of the 
deconfinement transition.

 In this note, we show
the abelian Polyakov loop is written by a product of monopole Dirac string
part and photon part and vanishing of the abelian Polyakov loop 
comes only from the monopole Dirac-string contributions 
not only in MA gauge but also in unitary gauges.

\section{Abelian Polyakov loop and monopole Dirac-string}

We adopt the usual $SU(2)$ and the $SU(3)$ Wilson actions.
To study gauge dependence, we consider three types of abelian 
projection.
The matrices to be diagonalized are $\sum_{\mu}
[U(s,\mu)\sigma_3 U^{\dagger}(s,\mu)+U^{\dagger}(s-\hat\mu,\mu)
\sigma_3 U(s-\hat\mu,\mu)]$ in MA gauge, Polyakov loop in Polyakov
gauge and plaquette in $F_{12}$ gauge.

After any gauge fixing is over, we can extract an abelian link
variable $u(s,\mu)$
and an angle variable $\theta_\mu (s)$ from it\cite{shiba}.

Now let us show that an abelian Polyakov loop operator 
of an external current $J_4 (s)$ 
\begin{eqnarray}
P = {\rm Re}[\exp\{i\sum_{s'=s}^{s+(N_4-1)\hat4} J_4 (s')\theta_4 (s')\}],
\label{apol}
\end{eqnarray}
is given  by a product of monopole and photon contributions. 

Using the definition of a plaquette variable 
$f_{\mu\nu}(s)= \partial_{\mu}\theta_{\nu}(s) - 
\partial_{\nu}\theta_{\mu}(s)$ 
, we get 
\begin{equation}
\theta_4 (s)= -\sum_{s'} D(s-s')[\partial'_{\nu}f_{\nu 4}(s')\\
        +\partial_4 (\partial'_{\nu}\theta_{\nu}(s'))], \label{t4}
\end{equation} 
where $D(s-s')$ is the lattice Coulomb propagator. 

Since $\partial'_4 J_4 (s) =0$, the second term in the right-hand side of 
(\ref{t4}) does not contribute to the abelian Polyakov loop (\ref{apol}).

\input epsf

\begin{figure}
\vspace{-1.8cm}
\epsfxsize=0.5\textwidth
\begin{center}
\leavevmode
\epsfbox{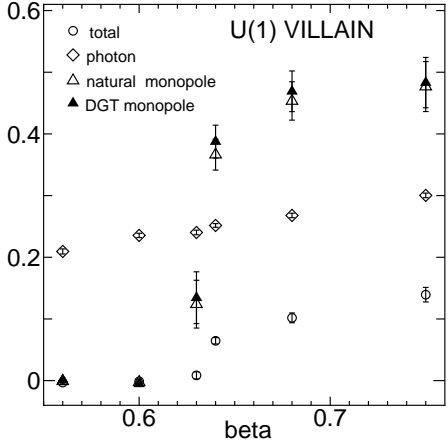}
\end{center}
\vspace{-2cm}
\caption{
Monopole Dirac string and photon contributions to Polyakov loops in the 
Villain model of compact QED.
}
\label{villain}
\end{figure}

The plaquette variable can be decomposed into three terms:

\begin{eqnarray}
\lefteqn{f_{\mu\nu}(s)  = -\epsilon_{\mu\nu\alpha\beta}
               \partial'_{\alpha}C_{\beta}(s)} \nonumber\\
 &    &\qquad    +\partial_{\mu}\bar{\theta}_{\nu}(s)
            -\partial_{\nu}\bar{\theta}_{\mu}(s) + 2\pi n_{\mu\nu}(s)
\label{t5}
\end{eqnarray} 
which $n_{\mu\nu}(s)$, $C_{\mu}$ and $\bar{\theta}_{\mu}$ are 
an integer-valued Dirac strings \cite{degrand},
a dual potential and a photon terms, respectively.
The dual potential term has no contribution because of the 
antisymmetric tensor.
 Therefore we have $P={\rm Re}[P_1 \cdot P_2]$
where
\begin{eqnarray}
\lefteqn{P_1  =  \exp\{-i\sum_{s'=s}^{s+(N_4-1)\hat4} J_4 (s'')}\nonumber\\
 & &   \sum_{s''}D(s'-s")\partial'_{\nu}
   (\partial_{\nu}\bar{\theta}_4(s'')-
         \partial_4\bar{\theta}_{\nu}(s''))\},\\
\lefteqn{P_2  =  \exp\{-2\pi i\sum_{s'=s}^{s+(N_4-1)\hat4}J_4(s')}\nonumber\\
 & &\quad       \sum_{s''}D(s'-s'')\partial'_{\nu}n_{\nu 4}(s'')\}.
\end{eqnarray}
We observe separately the real parts $P_{p}$ and $P_{m}$ 
of the photon $P_1$ and the Dirac-string $P_2$, respectively.

\section{Measurements}

\subsection{The Villain form of QED}

We first check that the separation works well in Villain form of 
the compact $U(1)$\cite{villain}on a $8^4$ lattice. 
Since there are natural monopoles and DeGrand-Toussaint(DGT) monopoles 
in the Villain case of QED, we observe 
$P_m$ in terms of the two types of the monopoles.
The results are shown in Fig. \ref{villain}.

1)The monopole Dirac-string data vanish in the confinement phase, 
whereas the photon data 
remain finite and change gradually for all $\beta$. 
The characteristic features of the Polyakov loops are then 
due to the behaviors of the Dirac-string contributions alone.

2)Monopole Polyakov loops show more enhancement than the total ones 
for $\beta > \beta_c$. 

3)Both types of monopoles give almost the same results.

\subsection{The MA gauge  in $SU(2)$ and $SU(3)$ QCD}

\input epsf

\begin{figure}
\vspace{-1.5cm}
\epsfxsize=0.5\textwidth
\begin{center}
\leavevmode
\epsfbox{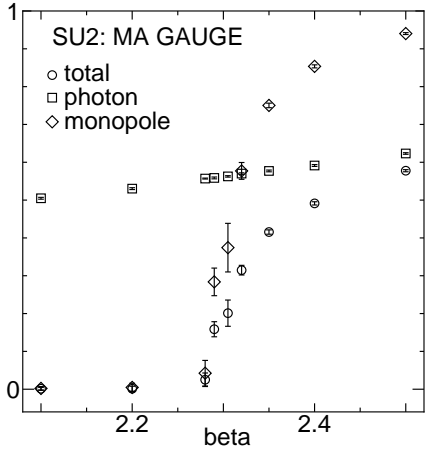}
\end{center}
\vspace{-2cm}
\caption{
Monopole Dirac string and photon contributions to Polyakov loops in the 
MA gauge in $SU(2)$ QCD.
}
\label{ma}
\end{figure}

The Monte-Carlo simulations were done in $SU(2)$ 
on $16^3\times 4$ lattice ($\beta =2.1\sim2.5$) in the MA and 
the unitary gauges. 
In $SU(3)$ QCD, we adopted $10^3 \times 2$ lattice ($\beta = 5.07\sim
5.12$).
 All measurements were done every 50 sweeps (40 sweeps in the $SU(3)$ case) 
after a thermalization of 2000 sweeps. We took 50 
configurations totally for measurements. 

1)We plot the $SU(2)$ data in the MA gauge in Fig.\ \ref{ma}.
The abelian Polyakov loops remain zero in the confinement phase, whereas 
they begin to rise for $\beta > \beta_c = 2.298$\cite{satz}.
The monopole Dirac string contribution is zero for $\beta < \beta_c$,
 whereas it begins to rise rapidly and it 
reaches $\sim 1.0$ for large $\beta$. On the other hand, the photon 
part has a finite contribution for both phases and it changes only slightly.
Characteristic behaviors of the abelian Polyakov loops are then explained 
by the Dirac-string part of monopoles alone. 

\begin{figure}
\vspace{-2cm}
\epsfxsize=0.5\textwidth
\begin{center}
\leavevmode
\epsfbox{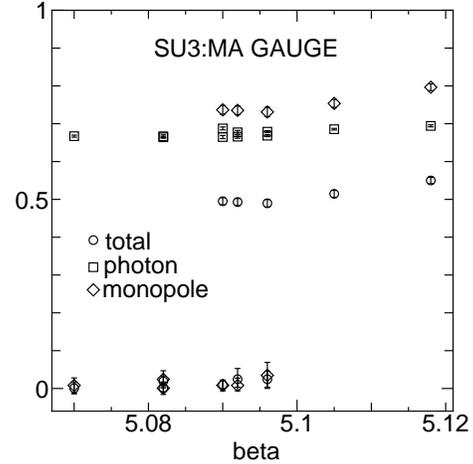}
\end{center}
\vspace{-2cm}
\caption{
Monopole Dirac string and photon contributions to Polyakov loops in the 
MA gauge in $SU(3)$ QCD.
}
\label{su3}
\end{figure}

2)The same results are obtained also in 
pure $SU(3)$ QCD in the MA gauge as shown in Fig.\ \ref{su3}. 
There is a clear hysteresis behavior showing the first order transition.

\input epsf

\subsection{The unitary gauges}

\input epsf

\begin{figure}
\vspace{-1.7cm}
\epsfxsize=0.5\textwidth
\begin{center}
\leavevmode
\epsfbox{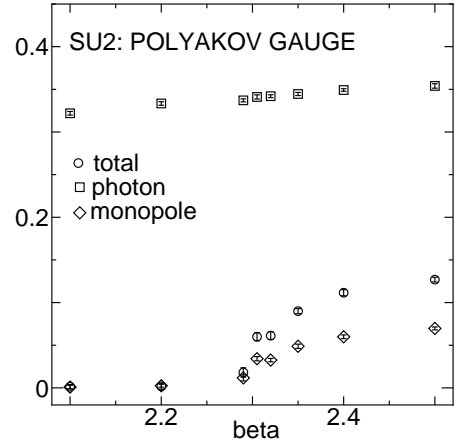}
\end{center}
\vspace{-2cm}
\caption{
Monopole Dirac string and photon contributions to Polyakov loops in the 
Polyakov gauge.
}
\label{pol}
\end{figure}

The data in the Polyakov gauge are plotted in Fig.\ \ref{pol}. 
It is very interesting to see that the abelian and the Dirac-string 
Polyakov loops are zero in the confinement phase, which 
suggests occurrence of flux squeezing in the unitary gauge, too. 
They show finite contribution above the critical 
temperature $\beta_c$. Photon contributions are finite and change 
 gradually in both phases. 
In the case of $F_{12}$ gauge, the results are very similar to these
in Polyakov gauge.

\section{Conclusion}
Our analyses done here show 
that abelian monopoles are responsible for 
confinement in $SU(2)$ and $SU(3)$ QCD, 
and condensation of the monopoles is the confinement mechanism. 
{\it These are the first phenomena suggesting gauge independent realization 
of the 'tHooft conjecture.}

\end{document}